# Spin and charge transport in material with spin-dependent conductivity

*V. Zayets*

*Spintronic Research Center, National Institute of Advanced Industrial Science and Technology (AIST), Umezono 1-1-1, Tsukuba, Ibaraki, Japan. E-mail: v.zayets@aist.go.jp*

*The spin and charge transport in materials with spin-dependent conductivity has been studied. It was shown that there is a charge accumulation along spin diffusion in a ferromagnetic metal, which causes a shortening of the spin diffusion length. It was shown that there is a substantial interaction between the drift and diffusion currents in semiconductors. The effects of gain/damping of a spin current by a charge current and the existence of a threshold spin current in a semiconductor were described. Because of the substantial magnitude, these new spintronics effect might be used for new designs of efficient spintronic devices. The influence of a spin drain on spin transport was discussed.*

1. Introduction

Spintronics is a new type of electronics that exploits the spin degree of freedom of an electron in addition to its charge. There are many expectations that in the near future, spintronics devices will be competitive with modern Si electronics devices. It is expected that spintronics devices will be faster, compacter and more energy-saving.

In the last decade there were significant advances in the field of spintronics. New effects, new functions and new devices have been explored. The spin polarized current was efficiently injected from a ferromagnetic metal into a non-magnetic metal and a semiconductor [1-6], the method of electrical detection of spin current was developed [6,7], the Spin Hall effect [8] and the inverse Spin Hall effect [9] were experimentally measured, the operation of a spin transistor[10] and tunnel spin transistor[11-13] were experimentally demonstrated.

However, the efficiency of spintronics devices is still low. For example, the operation of a spin transistor has been experimentally demonstrated [10], but the spin transistor operates only at low temperature and its output voltage is only a few micro volts. This voltage is too low for the transistor to be utilized in practical devices. This is the general tendency, except for devices based on a magnetic tunnel junction (MTJ) [14,15]. At present, most spintronic devices either operate at cryogenic temperatures or have a spin-dependent output signal, which is just a little above noise level [6,10]. It is important to improve the performance of spintronics devices, otherwise they have no chance of competing with modern Si-electronics devices. For example comparing the MOSFET transistor and the spin transistor, in the case of a MOSFET transistor the channel conductivity varies by several orders of magnitude between the on and off states, while in the spin transistor the voltage varies only by a few tenth of a percent. Here we study new spin-related effects, which might help to make spintronics devices more efficient and more competitive with Si electronics. We describe the effects of gain/damping of a spin current by a charge current and the existence of a threshold spin current in a semiconductor. An advantage of these new spin-related effects is that they may have significant magnitude. As shown below, due to these effects the conductivity of a semiconductor can be varied by a spin current within a broad range, which is comparable with the range of conductivity variation in the channel of a MOSFET transistor. Spintronics devices utilizing these effects may not only have new functions such as switching by spin current, but also be competitive in performance with Si devices.

In data processing circuits, a spin current can not be used as an input or output of a spintronic device (for example, a spin transistor), a charge current or an electrical voltage must be used. That implies that a spintronic device may operate efficiently only in the case when it utilizes an efficient conversion between spin current and charge current. For the last 20 years, spin transport in solids has been successfully described by the Valet-Fert spin diffusion equation [16]. The Valet-Fert equation describes spin diffusion from regions of larger spin accumulation towards regions of smaller spin accumulation. It does not include any term describing the interaction between a spin current and a charge current. Within the Valet-Fert theory spin current may interact with charge current only at a boundary between two materials. Therefore, the efficient conversion between the charge and spin currents and the efficient operation of the spintronics device may only be achieved in the case, when the device is utilizing interfaces with significantly spin-dependent resistance (for example, giant magnetic resistance (GMR) or tunnel magnetic resistance (TMR)). This requirement limits the range of possible designs of spintronics devices and it is technologically challenging to fabricate interfaces with high spin selectivity.

Recently, it has been noticed that the Valet-Fert equation is not sufficient for the description of all features of spin transport in ferromagnetic metals[17] and semiconductors[18]. In section 2 we will show that in the case of a material with spin-dependent conductivity the requirement of spin and charge conservation leads to the spin/charge transport equations, which are more complex than the Valet-Fert equation. These equations include terms that describe the interaction between charge and spin currents. As a result, the spin/charge transport equations do not exclude the possibility of conversion between spin and charge currents in the bulk of the material. This opens more possibilities for the design and fabrication of more efficient spintronics devices. In section 3 we will show that in the limit case when the conductivity of a material is spin-independent, the obtained spin/charge transport equations converge to the Valet-Fert spin transport equation. In section 4 we study the spin/charge transport in a ferromagnetic metal. Even though in the bulk of ferromagnetic metals there is no interaction between the dc spin and charge currents, we show that there is a charge accumulation along the spin diffusion. That leads to the interaction between ac spin and charge currents in the bulk of the ferromagnetic metal. In sections 5, 6 and 7 we study unique features of the spin transport in the semiconductors, which is greatly influenced by the interaction between charge and spin currents. In section 8 the influences of spin drain on the spin transport are explained.

**2. Spin and charge transport equations in materials with spin-dependent conductivity.**

In this section, the spin and charge transport equations for materials with spin-depending conductivity are derived from the spin and charge conservations laws. The spin/charge transport equations describe the diffusion of electrons from regions of higher spin/charge concentration to regions of lower spin/charge concentration. Also, they describe a drift of electrons under an applied electrical field.

A material may have a different conductivity $\sigma_\uparrow, \sigma_\downarrow$ for spin up and spin down electrons, which can be described as

$$\begin{pmatrix} \sigma_\uparrow \\ \sigma_\downarrow \end{pmatrix} = 0.5\sigma \left(1 + \beta \begin{pmatrix} 1 \\ -1 \end{pmatrix}\right) \quad (1)$$

where we define $\sigma$ as the effective conductivity and $\beta$ as the spin selectivity.

We assume that within each spin band the electrons are in thermo equilibrium. As result, the diffusion and drift of electrons in spin-up and spin-down bands can be described by spin dependent chemical potentials $\mu_\uparrow$ and $\mu_\downarrow$. Instead of $\mu_\uparrow$ and $\mu_\downarrow$, it is more convenient to use charge chemical potential $\mu_{charge}$ and the spin chemical potential $\mu_{spin}$ defined as

$$\begin{pmatrix} \mu_\uparrow \\ \mu_\downarrow \end{pmatrix} = \mu_{charge} \begin{pmatrix} 1 \\ 1 \end{pmatrix} + \mu_{spin} \begin{pmatrix} 1 \\ -1 \end{pmatrix} \quad (2)$$

The charge chemical potential $\mu_{charge}$ describes the drift and accumulation of charge and the spin chemical potential $\mu_{spin}$ describes diffusion and accumulation of the spin. A drift and diffusion of electrons in each spin band are described by Ohm's law

$$J_{\uparrow,\downarrow} = \sigma_{\uparrow,\downarrow} \nabla \mu_{\uparrow,\downarrow} \quad (4)$$

where $J_{\uparrow,\downarrow}$ and $\sigma_{\uparrow,\downarrow}$ are currents and conductivities for spin-up and spin-down electrons, respectively.

The flow of charge and flow of spin are described by charge and spin currents, respectively. They can be calculated as

$$J_{charge} = J_\uparrow + J_\downarrow \quad J_{spin} = J_\uparrow - J_\downarrow \quad (3)$$

Substituting (2) and (3) into (4), the charge and spin currents are calculated as

$$J_{charge} = \sigma \left(\nabla \mu_{charge} + \beta \cdot \nabla \mu_{spin}\right)$$
$$J_{spin} = \sigma \left(\nabla \mu_{spin} + \beta \cdot \nabla \mu_{charge}\right) \quad (4)$$

The conservation laws of charge and spin read

$$\nabla \cdot J_{charge} = \frac{\partial \rho_{charge}}{\partial t}$$
$$\nabla \cdot J_{spin} = \frac{\partial \rho_{spin}}{\partial t} \quad (5)$$

where $\rho_{charge}, \rho_{spin}$ are the charge and spin density, respectively.

In the static case there is no charge dissipation

$$\frac{\partial \rho_{charge}}{\partial t} = 0 \quad (6)$$

In the Appendix 1 it is shown that

$$\frac{\partial \rho_{spin}}{\partial t} = \sigma \frac{\mu_{spin}}{l_S^2} \quad (7)$$

where $l_s$ is defined as the intrinsic spin diffusion length.

Substituting (4), (6) and (7) into (5), the charge/spin transport equations are obtained as

$$\nabla \cdot \left[ \sigma \left( \nabla \mu_{charge} + \beta \cdot \nabla \mu_{spin} \right) \right] = 0$$

$$\nabla \cdot \left[ \sigma \left( \nabla \mu_{spin} + \beta \cdot \nabla \mu_{charge} \right) \right] = \sigma \frac{\mu_{spin}}{l_S^2} \quad (8)$$

It should be noticed that the equations (8) are derived only utilizing the spin/charge conservation laws and Ohm's law. Therefore, their validity extends over a wide range of material and structures.

In following sections we distinguish 3 types of materials, in which the spin transport substantially different. The first type of materials is non-magnetic metals, in which the conductivity is spin-independent. The second type of materials is ferromagnetic metals, in which the conductivity is spin-dependent and it is a constant throughout whole material. The conductivity of ferromagnetic metals does dependent on spin or/and charge accumulation within it. The third type of materials is semiconductors. The conductivity of semiconductors does dependent on spin or/and charge accumulation within it. The conductivity of semiconductors becomes spin-dependent in the presence of a spin-accumulation within it.

3. **Non-magnetic metals.**

A non-magnetic metal is defined here as a conductive material, the conductivity of which is independent on spin polarization ($\beta=0$).

In the case when $\beta=0$, Eqns.(8) can be simplified to

$$\nabla \cdot \left[ \sigma \nabla \mu_{charge} \right] = 0 \quad (9)$$

$$\nabla \cdot \left[ \sigma \nabla \mu_{spin} \right] = \sigma \frac{\mu_{spin}}{l_S^2} \quad (10)$$

Eqn. (9) describes the charge transport. Eqn. (10) is the Valet-Fert equation [16], which describes spin diffusion.

A solution of Eqn (9) describes a drift charge current $\vec{J}_{charge}$, which is drifted along an applied electrical field $\vec{E}$

$$\vec{J}_{charge} = \sigma \vec{E}$$
$$\vec{E} = \nabla \mu_{charge} \quad (11)$$

The Valet-Fert spin-diffusion equation (10) has a general solution

$$\mu_{spin}(\vec{r}) = \oint \mu_{spin}(\vec{s}) \cdot e^{-\frac{\vec{r} \cdot \vec{s}}{l_S}} d\vec{s} \quad (12)$$

where $\vec{S}$ is a unit vector directed toward the flow direction of the spin current. Along the diffusion direction the spin accumulation and the spin current exponentially decay.
As follows from Eqns. (9) and (10), in non-magnetic metals there is no interaction between the drift charge current and the diffusion spin current.

## 4. Ferromagnetic metals

In a ferromagnetic metal at Fermi energy there is a difference of the number of states for electrons, which spins are directed along and opposite to the metal magnetization direction. Since the conductivity of the metal is proportional to the number of carries participating in the transport and the number of spin-up and spin-down electrons is different, the conductivity of a ferromagnetic metal is different for spin-down and spin-up electrons. This difference in the conductivities is a constant throughout the bulk of the material and it is independent of the charge and the spin accumulation in the metal. As result, the spin selectivity β in the ferromagnetic metal is non-zero and it is a constant throughout the metal. A drift spin current and a charge accumulation along the spin diffusion current are features of spin transport in the ferromagnetic metals.
In the case when β is a constant, the spin/charge transport equations (8) can be simplified to

$$\nabla \cdot \left[ \sigma \nabla \mu_{charge} \right] = -\frac{\beta}{1-\beta^2} \cdot \sigma \frac{\mu_{spin}}{l_S^2} \quad (13)$$

$$\nabla \cdot \left[ \sigma \nabla \mu_{spin} \right] = \frac{1}{1-\beta^2} \cdot \sigma \frac{\mu_{spin}}{l_S^2} = \sigma \frac{\mu_{spin}}{l_{S,effective}^2} \quad (14)$$

where

$$l_{S,effective} = l_S \sqrt{1-\beta^2} \quad (15)$$

Equations (13),(14) are simpler than the general spin/charge equations (8), because the solution of Eqns. (13), (14) can be separated into two solutions. The first solution describes the drift current and the second solution describes the diffusion current. The same as in the case of non-magnetic metals, in ferromagnetic metals there is no interaction between the dc diffusion and drift currents. However, as it is shown below there is an interaction between the ac diffusion and drift currents in ferromagnetic metals, because of a charge accumulation along spin diffusion.
The first solution of Eqns. (13) (14) corresponding to the drift current is

$$\mu_{spin} = 0 \quad (16)$$

Eqn. (16) means that there is no spin accumulation along the drift current.
Substituting (16) into (4), the spin and charge components of the drift current are

$$\vec{J}_{charge} = \sigma \nabla \mu_{charge} = \sigma \cdot \vec{E}$$
$$\vec{J}_{spin} = \beta \cdot \sigma \nabla \mu_{charge} = \beta \cdot \sigma \cdot \vec{E} \quad (17)$$

The drift current has both a spin-current component and a charge-current component. In contrast to the diffusion spin current, the drift spin current does not decay with the propagation distance and there is no spin relaxation along the flow of the spin drift current.

The second solution of Eqns. (13) and (14) corresponding to a diffusion spin current is

$$\mu_{spin} \neq 0 \quad (18)$$

There is a spin accumulation along the flow of the diffusion spin current. The set of Eqns. (13),(14) has two unknowns: $\mu_{spin}$ and $\mu_{charge}$. $\mu_{spin}$ can be found from Eqn. (14). Eqn. (14) is the same as the Valet-Fert equation (10). As result, the solution for $\mu_{spin}$ is described by (12). However, the effective spin diffusion length (15) should be substituted in place of the intrinsic spin diffusion length (7). Knowing $\mu_{spin}$, the $\mu_{charge}$ is found from (13) as

$$\mu_{charge} = -\beta \cdot \mu_{spin} \quad (19)$$

The properties of the spin diffusion current are similar in ferromagnetic and non-magnetic metals. However, the effective spin diffusion length in ferromagnetic metals is shorter, because in a ferromagnetic metal there is a charge accumulation along the flow direction of the spin current and the electrical field induced by this charge accumulation is slowing the spin diffusion.

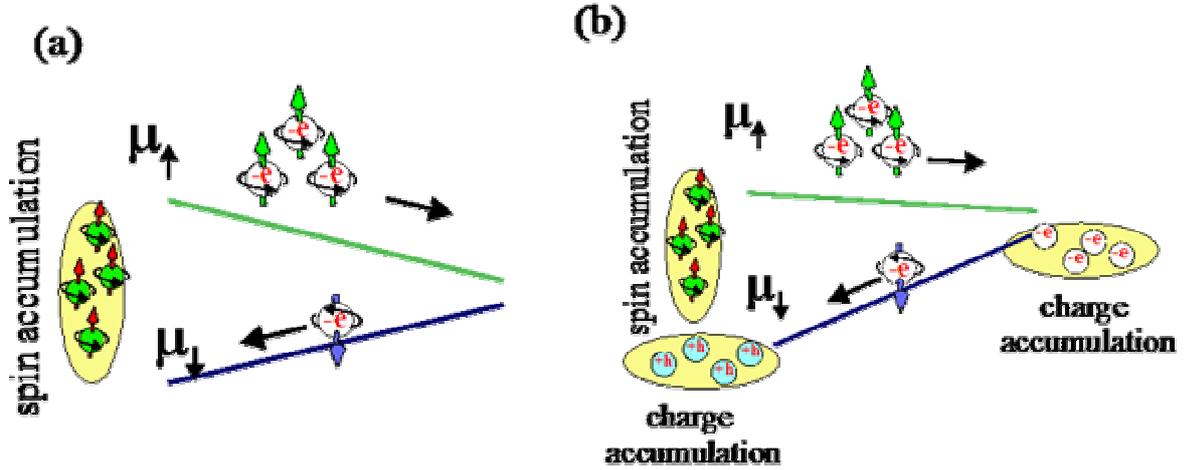

*Fig.1. Physical mechanism for charge accumulation along the spin diffusion current in ferromagnetic metal. Green and blue lines represent chemical potential for spin-up and spin-down electrons, respectively. (a) Spin-up electrons are accumulated at left side and there is spin diffusion towards right side. Diffusion of spin-up and spin-down electrons is unbalance, charge is accumulating and transport is not in equilibrium. (b) The charge is accumulated, diffusion of spin-up and spin-down electrons is in balance and transport is equilibrium*

The charge accumulation along the diffusion spin current is calculated from Gauss's law utilizing (19) as

$$\rho_{charge} = \nabla \cdot \left[\varepsilon \vec{E}\right] = \nabla \cdot \left[\varepsilon \nabla \mu_{charge}\right] = \nabla \cdot \left[\varepsilon \nabla \left(\beta \cdot \mu_{spin}\right)\right] \quad (20)$$

where $\rho_{charge}$ is the accumulated charge density and $\varepsilon$ is the permittivity of the metal.

For example, in the case when spin diffusion is along the x-direction, the spin chemical potential and charge accumulation are

$$\mu_{spin} = \mu_{spin0} e^{-\frac{x}{l_{S,effective}}} \quad (21)$$

$$\rho_{charge} = -\beta \cdot \varepsilon \frac{\mu_{spin0}}{l_{S,effective}^2} e^{-\frac{x}{l_{S,effective}}} \quad (22)$$

Figure 1 explains the physical mechanism for charge accumulation along the spin diffusion current in a ferromagnetic metal. The magnetization direction of the ferromagnetic metal is up. As result, the density of states for spin-up electrons is larger than the density of states for spin-down electrons. We assumed that there is a spin accumulation of spin-up electrons at left side and there is a spin diffusion towards the right side. Since spin-up electrons are accumulated and the accumulation decays towards right side, the chemical potential for spin up electrons $\mu_\uparrow$ is higher than for spin down electrons $\mu_\downarrow$ and the differences between the chemical potentials $\mu_\uparrow$ and $\mu_\downarrow$ decreases towards the right side. $\mu_\uparrow$ is decreasing and $\mu_\downarrow$ is increasing. That means that the spin-up electrons diffuse from left side to right side and spin-down electrons diffuse in opposite direction. In the case when there is no charge accumulation (Fig. 1 (a)), the slopes of $\mu_\uparrow$ and $\mu_\downarrow$ are of the same magnitude and of opposite signs. Since the density of states for spin-up electrons is greater than for spin-down electrons, more electrons diffuse toward the right side than toward the left side. Because of the unbalance of electrons diffusing in opposite direction, there is a charge current and the charge is accumulating in the bulk of the metal. The charge accumulation induces an electrical field. The direction of the electrical field is such that it reduces the slope of the chemical potential for spin up electrons and increases the slope for spin-down electrons (Fig. 1(b)). As a result, the diffusion of spin-down electrons increases and the diffusion of spin-up electrons decreases. The charge accumulation proceeds until the diffusion of spin-down electrons from right to left will be equal to the diffusion of spin-up electrons from left to right.

As described by Eqn.(22), the charge accumulation is larger in a material with a shorter spin diffusion length. The materials of largest spin selectivity $\beta \sim 1$ have the shortest effective spin diffusion length (See Eqn. (15)) and the largest charge accumulation. A semi-metal is a material, in which the density of states for one spin band is zero near the Fermi energy and electrons of only one spin polarization are able to participate in the transport. In the case of semi-metals, the spin selectivity β=1 and the spin diffusion length becomes zero (See Eqn. (15) ) meaning that in semi-metals the spin diffusion can not occur. This fact can be understood from Fig.1. When there is a spin accumulation in the semi-metal, the spin starts to diffuse (for example of Fig.1, from left to right), immediately causing a charge accumulation and an electrical field in the direction opposite to the diffusion, so that an equal amount of electrons is drifted back. As for example of Fig.1, the number of electrons that diffuse from left to right due to spin accumulation equals to number of electrons that diffuse from right to left due to charge accumulation. Therefore, the spin accumulation in a semi-metal does not spread in space.

As shown below, the absence of spin diffusion at β=1 is the reason for a threshold spin current in a semiconductor. The diffusion spin current in a semiconductor can not exceed the threshold spin current.

As an example, in the ferromagnetic metal (β=0.5 ε=10) the spin accumulation of $\mu_{spin}$=1 mV is accompanied by an accumulated electron density of 2.8E11 cm-3 and 2.8E15 cm-3 in the cases of the effective spin diffusion length of 1 um and 10 nm, respectively. In this example, the accumulated electron density is several orders of magnitude smaller than the density of conductive electrons in the metal.

As was shown above, in a ferromagnetic metal the drift and spin currents flow independently and there is no conversion between them. It is important to clarify the conditions when the drift spin current may be converted into the spin diffusion current. Eqns. (14), (15) do not describe any conversion between drift and diffusion spin currents. These equations were derived from the general spin/charge transport equations (8) only under one condition: $\nabla \beta = 0$. Therefore, it can be concluded that the conversion of drift spin current into diffusion spin is only possible when

$$\nabla \beta \neq 0 \quad (17)$$

The condition (17) may be satisfied at the boundary between two materials. For example, the conversion of a drift spin current in a ferromagnetic metal into a diffusive spin in a non-magnetic metal is called spin injection. The effective spin injection is important for the operation of spintronics devices.
In the following chapter, we will show that condition (17) is satisfied in the volume of semiconductors. Therefore, in the volume of a semiconductor, the spin drift current is continuously converted into the diffusion spin current.

5. **Semiconductors**

In a semiconductor the spin selectivity β is not constant throughout the bulk of the material. At each point it is a function of the magnitude of the spin accumulation there. Because of this condition, there are several unique features of spin transport in the semiconductors such as: a spin conversion between a drift and diffusion currents; an existence of a threshold spin current; and a gain/damping of the spin current by a charge current.
In the following, we will derive a set of equations, which describe the spin and charge transport in non-degenerate semiconductors. In the case of a non-degenerate n-type semiconductor, the number of electron in conduction band is

$$n = N_c e^{\frac{E_F - E_c}{kT}} \quad (18)$$

where $N_c$ is the effective density of states in the conduction band, $E_F$ is the Fermi energy, $E_c$ is the energy of the bottom of the conduction band.
In a semiconductor both the charge and spin can be accumulated. In the case when there is a charge accumulation, the number of electron in the conduction band is

$$N_c e^{\frac{E_F - E_C}{kT}} = N_{Doping} + n_{accumul} \quad (19),$$

where $N_{Doping}, n_{accumul}$ are the doping concentration and the concentration of accumulated electrons, respectively.

In the case when additionally there is a spin accumulation, the number of spin up and spin-down electrons in the conduction band is

$$\begin{pmatrix} n_\uparrow \\ n_\downarrow \end{pmatrix} = N_c e^{\frac{E_F - E_C}{kT}} \begin{pmatrix} e^{\frac{\mu_{spin}}{kT}} \\ e^{-\frac{\mu_{spin}}{kT}} \end{pmatrix} \quad (20)$$

Noticing that in non-degenerated semiconductor the mobility only weakly depends on charge and spin accumulations, the conductivity for spin-up and spin-down electrons can be calculated as

$$\begin{pmatrix} \sigma_\uparrow \\ \sigma_\downarrow \end{pmatrix} = e \cdot (N_{Doping} + n_{accumul}) \cdot mobility \cdot \begin{pmatrix} e^{\frac{\mu_{spin}}{kT}} \\ e^{-\frac{\mu_{spin}}{kT}} \end{pmatrix} \quad (21)$$

The effective conductivity σ and spin selectivity β in a semiconductor are calculated from (21) as

$$\sigma = \sigma_\uparrow + \sigma_\downarrow = e \cdot (N_{Doping} + n_{accumul}) \cdot mobility \cdot 2 \cdot \cosh\left(\frac{\mu_{spin}}{kT}\right) \quad (22)$$

$$\beta = \frac{\sigma_\uparrow - \sigma_\downarrow}{\sigma_\uparrow + \sigma_\downarrow} = \tanh\left(\frac{\mu_{spin}}{kT}\right) \quad (23)$$

Eqns. (22),(23), (8) and Gauss's law form a full set of equations describing the charge and spin transport in semiconductors. Therefore, in semiconductors the spin and charge transport are described by the following set of 5 equations in 5 unknowns ($\mu_{charge}$, $\mu_{spin}$, σ, β, $n_{accumul}$):

$$\nabla \cdot \left[\sigma(\nabla \mu_{charge} + \beta \cdot \nabla \mu_{spin})\right] = 0 \quad (24a)$$

$$\nabla \cdot \left[\sigma(\nabla \mu_{spin} + \beta \cdot \nabla \mu_{charge})\right] = \sigma \frac{\mu_{spin}}{l_S^2} \quad (24b)$$

$$\rho_{charge} = e \cdot n_{accam} = \nabla \cdot \left[\varepsilon \vec{E}\right] = \nabla \cdot \left[\varepsilon \nabla \mu_{charge}\right] \quad (24c)$$

$$\sigma = e \cdot mobility \cdot (N_{doping} + n_{accam}) \cdot 2 \cdot \cosh\left(\frac{\mu_{spin}}{kT}\right) \quad (24d)$$

$$\beta = \tanh\left(\frac{\mu_{spin}}{kT}\right) \quad (24e)$$

The equations of the set (24) are non-linear and in a general case they should be solved numerically.

## 6. Threshold diffusion spin current

A unique feature of spin transport in a semiconductor is the existence of a threshold spin current, above which the diffusion spin current is unable to flow and spin diffusion is stopped. In the case when an input spin current is slightly below the threshold current, the region of significant spin and charge accumulations is formed close to the input of the spin current. Only a small amount of spin current diffuses out of this region. The spin and charge accumulation in this region are substantial and the region's transport properties may be controlled within wide range by varying the input spin current. That makes this effect attractive for new designs of effective spintronics devices.

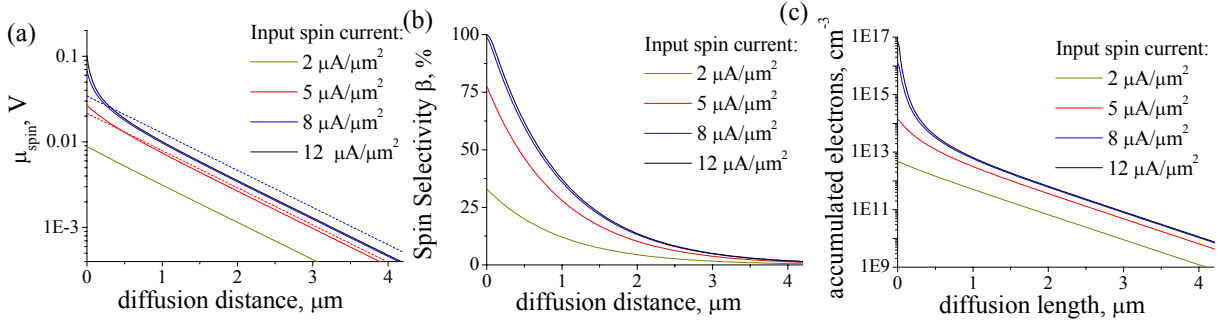

*Fig. 2 Threshold spin current in the semiconductor. (a) Spin chemical potential, (b) spin selectivity and (c) density of accumulated electrons for diffusive spin current propagating along n-Si bar (doping concentration $10^{16}$ cm$^{-3}$). The region of high spin/charge accumulation is from 0 to ~0.2 μm. The dash lines in (a) shows the case when spin current flow in a non-magnetic metal having the same spin diffusion length and the same conductivity. In cases (a) and (c), the scale of y-axis is logarithmic.*

To demonstrate this effect, we have solved numerically the set of equations (24) by the Finite Difference Method (FDM) for an n-Si bar (mobility of 1450 cm$^2$/V/s, spin diffusion length of 1 μm, doping concentration of $10^{16}$ cm$^{-3}$). The Neumann boundary conditions of zero charge current and a given value of spin current were used as the boundary conditions at the left-side of the bar. Open-boundary conditions for the spin current and a zero value for the charge chemical potential were used as the boundary conditions at the right side of the bar. Figure 2 shows the calculated spin chemical potential $\mu_{spin}$, the spin selectivity $\beta$ and the accumulated electrons density. In the case when the input current is below 2 μA/μm$^2$ the spin selectivity $\beta$ and the charge accumulation are small and the spin diffusion in the semiconductor is almost the same as the spin diffusion in a non-magnetic metal. The decay of the spin current $J_{spin}$ and the spin chemical potential $\mu_{spin}$ follow the exponential law with the effective spin diffusion length equal to intrinsic spin diffusion length in the semiconductor. This corresponds to the straight dark-yellow line of Fig.2 (a). However, for larger magnitudes of the input spin current, the spin diffusion in semiconductors becomes different from the spin diffusion in non-magnetic metals and the region of high spin/charge accumulation is formed. The decay of spin chemical potential $\mu_{spin}$ deviates from the exponential law (red, blue, black lines of Fig. 2(a)). Inside the accumulation region, the spin charge accumulation is larger compared to the case of a non-magnetic metal. Outside this region the spin accumulation is smaller. As can be seen from Fig.2(c), the large spin accumulation is accompanied by a large charge accumulation.

The physical mechanism for the formation of a region of high spin/charge accumulation is explained as follows. As can be seen from Fig. 2(b), in case of a small input spin current, the spin selectivity $\beta$ in a semiconductor is small and it has a little influence on the transport. In the case when the input spin current exceeds 5 μA/μm$^2$, the spin selectivity $\beta$ approaches 100 % in the region close to the input. As has been shown in section 4, in the case when the spin selectivity β approaches 100 % the spin transport drastically changes. The effective spin length shortens, significant charge is accumulated and eventually spin diffusion stops. In a semiconductor, the spin selectivity β increases with the increase of the magnitude of the spin current. Inside the region near the input where β approaches 100%, the spin diffusion length significantly shortens, meaning that there spin diffusion practically stops. Only a small amount of spin current leaks from this region. Since spin current does not diffuse out of this region, more spin is accumulated there, causing an even larger β and an even larger spin accumulation. Therefore, the process of spin/charge accumulation is avalanche-like. As can be seen from Fig.2 (a), when the input spin current increases further from 8 µA/µm² to 12 µA/µm², the amount of the spin current leaked from the accumulation region almost does not change. This means the spin current can not pass into the Si bar and so the majority of the spin current decays inside the spin/charge accumulation region.

Even though the explained effect limits the amount of spin current, which could pass through the semiconductor, the effect may be used in some applications. For example, the electron accumulation in the spin/charge accumulation region may significantly exceed the doping concentration in the semiconductor. This leads to a substantial increase of the conductivity of this region, meaning that a spin current may switch a semiconductor from a weakly-conductive state into a larger-conductivity state.

It should be noticed that the threshold spin current decreases with the decrease of temperature and the doping concentration.

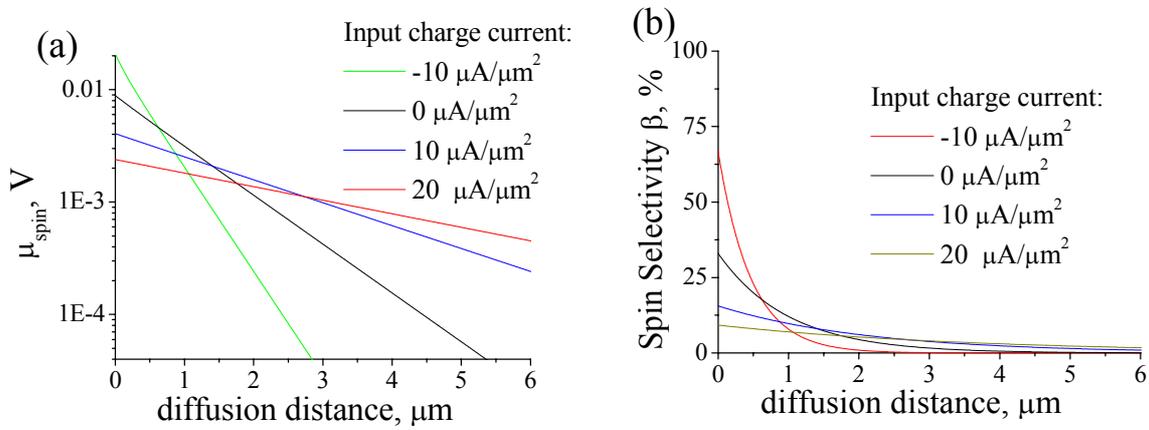

Fig. 3 The gain/damping of diffusive spin current by drift charge current. Input spin current is 2 µA/µm².
(a) Spin chemical potential $\mu_{spin}$. (b) Spin selectivity β

## 7. Gain/damping of diffusion spin current by charge current

In the previous section, the properties of the diffusion spin current in a semiconductor have been studied in the absence of any drift current. When drift and diffusion currents flows along the same direction, there is an interaction between them. One consequence of this interaction is a gain/damping of the spin current by the charge current.

As an example, the spin and charge transport in the n-Si bar (mobility of 1450 cm$^2$/V/s, spin diffusion length of 1 μm and doping concentration of $10^{16}$ cm$^{-3}$) was calculated by numerically solving Eqns (24) by the FDM. The Neumann boundary conditions for the given values of the spin and charge currents were used as the boundary conditions. The input spin current was 2 μA/μm$^2$.

Figure 3(a) shows the calculated spin chemical potential $\mu_{spin}$ for different values of the input charge current. For any input charge current, log($\mu_{spin}$) is a straight line. This means that despite the fact that the spin transport in semiconductors is described by a set of non-linear equations and there is a significant interaction between the charge and spin currents, still the decay of the spin chemical potential $\mu_{spin}$ is practically exponential, meaning that the spin transport can be described by an effective spin diffusion length. As can be seen from Fig. 3 (a), the effective spin diffusion length increases when the charge and spin currents flow in the same direction, and the effective spin diffusion length decreases, when the charge and spin currents flow in the opposite directions. When the charge current flows along the spin current, the spin current gains from the charge current so that the spin relaxation becomes slower and the spin length becomes longer. For the opposite direction of the charge current, the spin current is damped by the charge current so that the spin relaxation becomes faster and the spin length becomes shorter. It was calculated that the effective spin length does not depend on the magnitude of the input spin current.

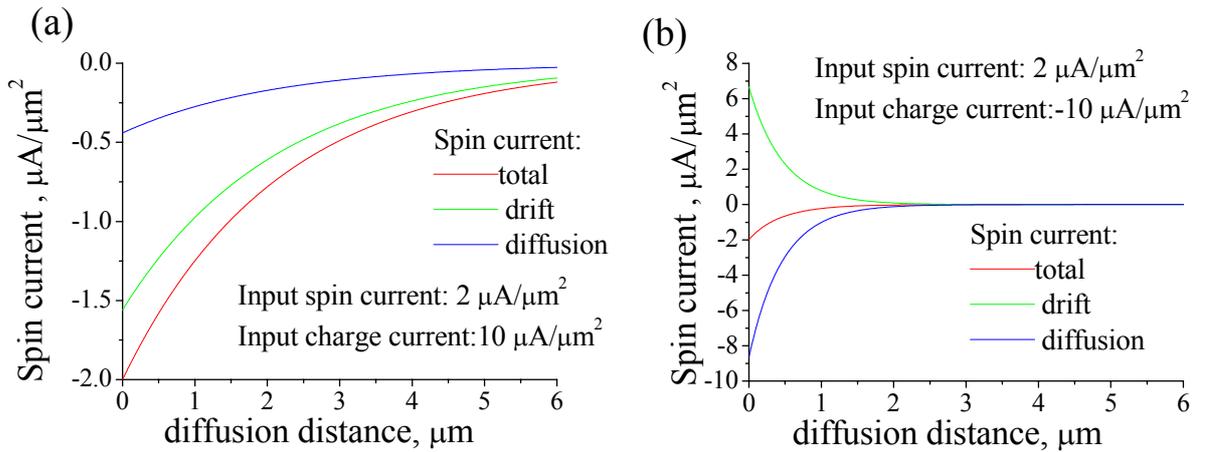

*Figure 4. The total, drift and diffusion spin currents in case when the drift current flows (a) along diffusion current (2) opposite to diffusion current*

The physical mechanism for the gain/damping of the spin current by the charge current is described as follows. The spin selectivity β in semiconductors is a function (23) of the spin accumulation. As was explained in section 3, in materials with non-zero spin selectivity β the drift current is spin-polarized and the magnitude of the spin-current component of the drift current is linearly proportional to β (See Eqn.(17)). This means that in the semiconductor the drift current is spin-polarized, only when it flows together with the diffusion spin current. Along the diffusion direction of spin current, the spin accumulation exponentially decays and consequently the spin selectivity β decays as well. Figure 3 (b) shows the spin selectivity β for different values of the input charge current. Since β decreases, the spin component of the drift current decreases as well. Because of the spin conservation law, the spins are converted from a drift current to a diffusion current. When the polarity of the converted spins is the same as the spin polarity of diffusion current, the diffusion current gains the spins and the decay of diffusion current becomes slower. When the polarity of the conversion is different, the diffusion current is damped and the decay of diffusion current becomes faster.

Figure 4(a) shows the diffusion spin current, the spin-current component of the drift current and their sum for the case when the drift current flows along the direction of the diffusion current. The polarity of the drift and diffusion spin currents are the same and the magnitude of each current is smaller than the magnitude of total spin current. Figure 4 (b) shows the diffusion spin current, the spin-current component of drift currents and their sum for the case when the drift and diffusion currents flow in opposite directions. The polarities of the drift and diffusion spin currents are opposite and the magnitudes of both diffusion and spin currents are substantially larger than the magnitude of the total spin current.

It should be noticed that spin relaxation affects only the diffusion current. There is no spin relaxation for the drift spin current. The drift spin current is loosing its spin current component only due the conversion of spins into the diffusion current. The spin relaxation is weakest for a given total spin current when the drift spin current is largest and diffusion spin current is smallest. For the same total input spin current, the magnitude of the diffusion spin current is significantly smaller and as consequence the spin relaxation is weaker in the case, when the flow directions of the charge and spin currents are the same, than in the case when the directions are opposite (Fig.4). Therefore, since spin relaxation is proportional to the magnitude of the diffusion spin current, the spin relaxation in semiconductors may be modulated by a charge current.

Even though spin/charge transport in the semiconductors is described by a set of non-linear equations, it is possible to find the analytical expression for the effective spin diffusion length. It is necessary to use several approximations. However, in many cases the analytical expression for the effective diffusion length is a good approximation, which describes well the spin transport in semiconductors. Substituting (4) into (24b) gives

$$(1-\beta^2)\nabla\cdot(\sigma\nabla\mu_{spin}) + \nabla\beta\cdot J_{charge} - 2\cdot\nabla\beta\cdot\beta\cdot\sigma\nabla\mu_{spin} = \sigma\frac{\mu_{spin}}{l_S^2} \quad (25)$$

In order to solve Eqn. (25) the following approximations are used. The spin current is assumed to be sufficiently below the threshold current. It is assumed that there are no regions of large spin or charge accumulation. The convection term $\nabla\beta\cdot\beta\cdot\sigma\nabla\mu_{spin}$ is ignored. The conductivity and charge current are assumed to be a constant throughout the material. Therefore, Eqn. (25) is simplified to

$$\nabla\cdot(\nabla\mu_{spin}) + \nabla\beta\cdot\frac{J_{charge}}{\sigma} = \frac{\mu_{spin}}{l_S^2} \quad (26)$$

Differentiating Eqn (24e) gives

$$\nabla\cdot\beta = \left[\mathrm{sech}\left(\frac{\mu_s}{kT}\right)\right]^2 \frac{1}{kT}\nabla\cdot\mu_s \quad (27)$$

In the case of a small spin accumulation when $\frac{\mu_s}{kT} \ll 1$, Eqn.(27) is simplified to

$$\nabla \cdot \beta = \frac{1}{kT} \nabla \cdot \mu_s \quad (28)$$

Substituting (28) into (26) gives

$$\nabla \cdot (\nabla \mu_{spin}) + \frac{1}{kT\sigma} \nabla \cdot \mu_{spin} \cdot J_{charge} = \frac{\mu_{spin}}{l_S^2} \quad (29)$$

The solution of Eqn. (29) is the same as the solution (14) of the Valet-Fert equation (10), but instead of the intrinsic spin diffusion length $l_S$ the effective spin diffusion length $l_{s,eff}$ should be used

$$l_{s,eff} = \frac{l_s}{\sqrt{1 + \left(\frac{J_{charge} l_S}{2kT\sigma}\right)^2} + \frac{J_{charge} l_S}{2kT\sigma}} \quad (30)$$

Figure 5 shows the effective spin diffusion length $l_{S,eff}$ in n-Si ($N_d=10^{16}$ cm$^{-3}$). The black line is obtained by fitting the FDM solutions of Fig. (3). The red line is calculated from Eqn. (30). The difference between the lines is less than 1 %. Therefore, despite the somewhat rough approximations used, Eqn. (30) well describes the interaction between the spin and charge currents in semiconductors in the majority of cases.

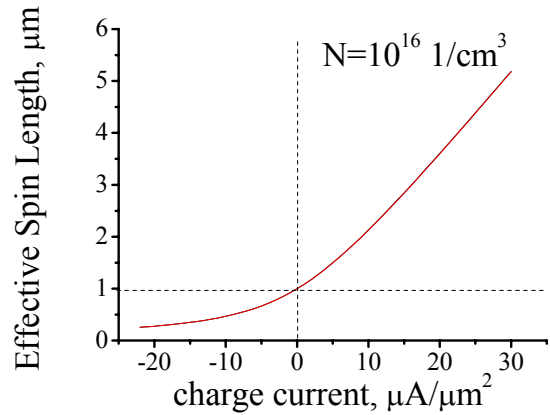

Fig.5 Effective spin diffusion length in n-Si ($N_d=10^{16}$ cm$^{-3}$). Black line is obtained by fitting FDM solutions of Fig. (3). Red line is calculated from Eqn. (30)

It should be noticed that the charge current can modulate the threshold spin current. As was described in the previous chapter, the threshold spin current is the maximum diffusion spin current, which can flow in a semiconductor. In the case when the input spin current is fixed and the input charge current is varied, the magnitude of the diffusion spin current may change significantly (See Fig.4). Therefore, the charge current may turn the diffusion spin current to or out of the threshold conditions, when the spin diffusion is stopped and a region of high spin and charge accumulation is formed. As result, the charge current may switch the spin diffusion on and off.

## 8. Spin drain effect

In contrast to the charge current, the flow of the spin current does not require a drain. However, the spin drain plays an important a role for spin transport. In the following we will demonstrate that a spin drain substantially modifies spin diffusion and may change the effective spin diffusion length.

Even though, a spin and a charge are undivided features of an electron, there is a significant difference between the flows of charge and spin in a solid. In contrast to the charge current, which always flows from source to drain, a flow of spin current does not require a drain. After having been emitted from a source, the spin current propagates in material until it decays.

Both a charge source and a charge drain are required in order for the charge current to flow. The charge is drifted by an electrical field from the charge source towards the charge drain. In the absence of the charge drain, the charge current does not flow. For example, in the imaginary case when there is no charge drain, but there a charge flow, the charge will be accumulated at some point. The accumulated charge will induce a long-range electrical field, which is in the opposite direction to the charge current. Therefore, the charge flow will stop.

In contrast, only a spin source is necessary for the spin current to flow. The flow of the spin current does not require a spin drain. A spin accumulation does not induce any sufficiently-strong long-range field, which may stop the spin diffusion. As a spin accumulation is created at the spin source, it diffuses form the regions of larger spin concentration towards the regions of smaller spin concentration.

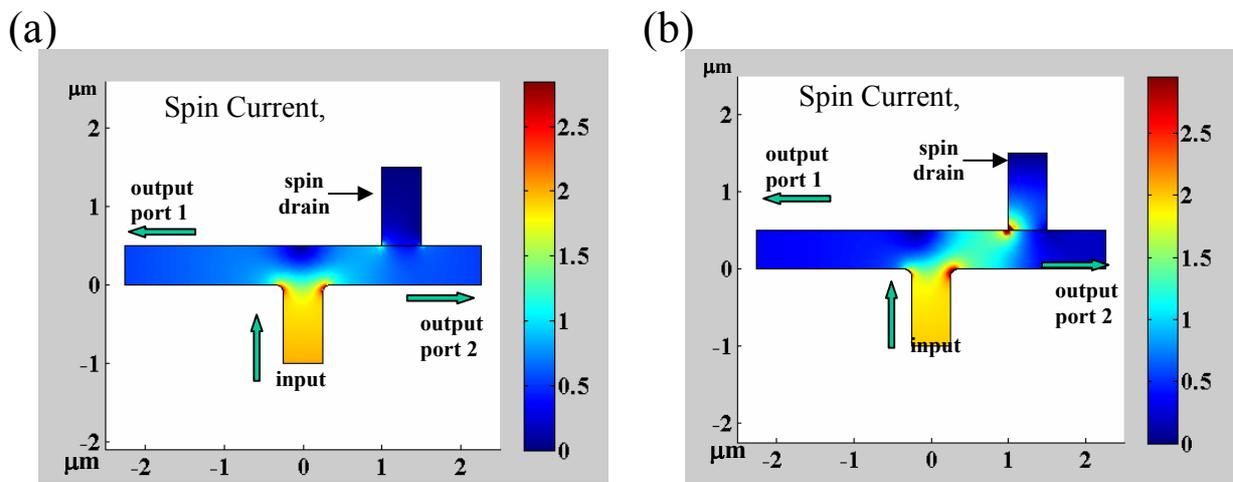

*Fig.6 Distribution of spin current in T-splitter with spin drain. (a) conductivities of spin drain and T-splitter are the same. (b) conductivity of spin drain is 20 times larger than conductivity of T-splitter*

Even though the spin current does not require a spin drain, the spin drain plays an important role in the spin transport. Any conductive material, in which the spin current rapidly decays, may be considered as a spin drain. A spin drain, which consists of a material of smaller conductivity and of longer spin diffusion length, has the stronger influence on spin transport.

To demonstrate the influence of the spin drain on spin transport, we calculated the spin diffusion in 3 example structures. The first example will demonstrate that a spin drain attracts spins. The second example will demonstrate that the spin drain affects the spin diffusion length in a material. The third example demonstrates the influence of the sample shape on the spin drain effect. All structures consist of a metallic wire and a spin drain. The materials of the wire and the drain were non-magnetic metals. The obtained results are valid also in the case of ferromagnetic metals and semiconductors, but the effective length (15) or (30) should be used instead of the intrinsic spin diffusion length (7). The solution was calculated by solving the Gerber-Fert diffusion equation (10) by the Finite Element Method. The conductivity and thickness of the wires were 3.774e7 S/m and 1 um, respectively. The spin drain is a metallic wire with larger conductivity. The spin drain effectively attracts spin current only when its conductivity is significantly larger than the conductivity of other wires in the structure. In order to clarify the influence of the spin drain, the spin diffusion was compared between the case, when the drain conductivity is smaller and the influence of spin drain is not significant, and the case, when the drain conductivity is larger and the influence of spin drain is significant.

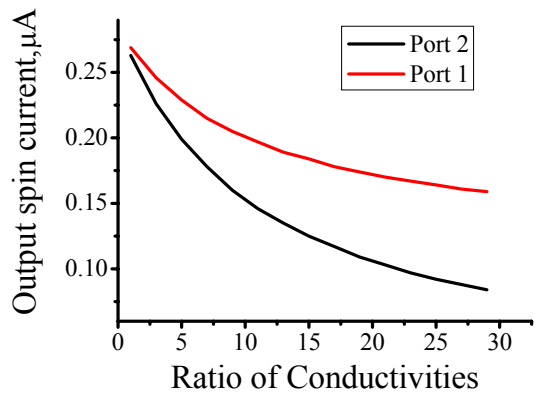

*Fig. 7 Output spin current at Ports 1 and 2 as function of ration of conductivity of spin drain to conductivity of T-splitter.*

For the first example we have calculated the spin diffusion in a T-splitter, which consist of an input wire and two output wires (Fig.6). There is a spin drain connected near the output of the second wire. The spin diffusion length in the wires of the splitter and in the drain is 5 μm. Figure 6(a) shows the distribution of the spin current in the T-splitter in the case when the conductivity of the spin drain is the same as that of the wires of the T-splitter. In this case the spin drain has little influence on the spin transport and almost equal amounts of spin current split between the two ports. The figure 6(b) shows the case when the conductivity of the spin drain is 20 times larger than the conductivity of the wires of the T-splitter. The spin current is mostly directed into the spin drain, because the spin drain attracts the spin current.

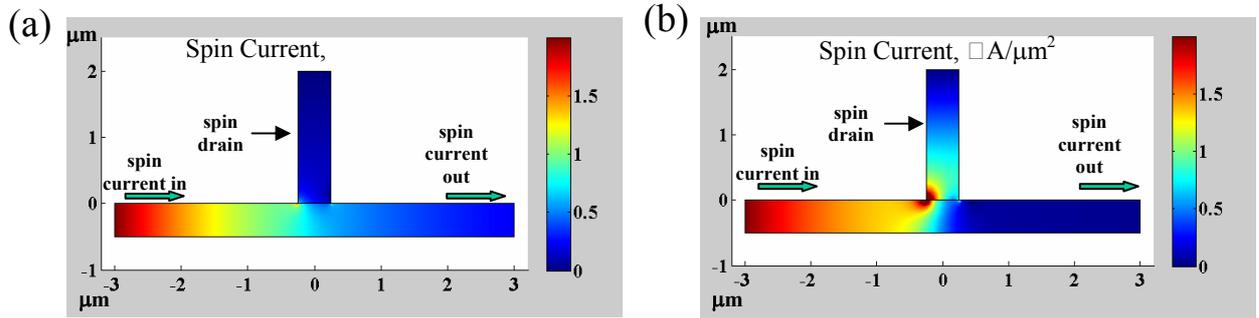

*Fig. 8. Spin current distribution in straight wire connected to spin drain. The spin diffusion length of the wire and the spin drain is 3 um. Ratio of conductivities of spin drain to wire is (a) 0.4 (b) 40*

Figure 7 shows the calculated output spin currents at Ports 1 and 2 as a function of the ratio of the conductivity of the spin drain to the conductivity of the wire of the T-splitter. In the case when the conductivity of the spin drain is small, the influence of the spin drain on the spin transport is weak and the spin current is split almost equally between ports 1 and 2. As the conductivity of the spin drain decreases, more spin current decays inside the spin drain and the output spin current from ports 1 and 2 rapidly decreases. Since port 2 is closer to the spin drain, the reduction of the spin current at this port is more rapid. However, the influence of the spin drain on the spin current in Port 1 is still significant despite the long distance between port 1 and the spin drain.

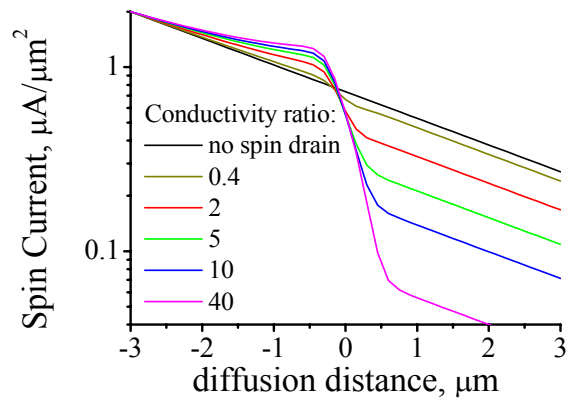

*Fig.9 Magnitude of spin current along diffusion in the wire for different ratios of conductivity of spin drain to the conductivity of the wire*

The second example demonstrates the influence of the spin drain on the effective spin diffusion length in a metallic wire. The structure consists of a straight metallic wire with the spin drain connected in the middle. Figure 8 shows the distribution of the spin current in this structure. In the case when the conductivity of the spin drain is smaller than the conductivity of the wire (Fig. 8(a)), the spin current mainly flows in the wire and only a little of the current enters the spin drain. In the case when the conductivity of the spin drain is larger than the conductivity of the wire (Fig. 8(b)), all spin current flows into the spin drain and only a little of current leaks into the wire. Figure 9 shows the magnitude of the spin current at center of the wire. The influence of the spin drain on the spin diffusion length may be notices from this figure. When the spin current has passed the spin drain, the spin drain does not affect the spin diffusion and the decay slope of the spin current is the same as in the case without the spin drain. However, before the spin drain the decay of the spin current is slower than in the case without the spin drain. That means the effective spin diffusion length becomes longer due to the spin drain. Since the spin current is attracted by spin drain, spin diffusion is faster and as result the spin current dissipates over a longer distance. This is a reason for the shortening of the spin diffusion length in the presence of a spin drain.

The spin drain is just a metallic wire having a sufficiently high conductivity and a sufficiently short spin diffusion length, meaning, that any conductive wire may be considered as a spin drain. This implies that the shape of the wire might affect spin diffusion. The third example demonstrates the influence of the sample shape on the spin drain effect. The structure consists of two connected straight wires of different widths. Figure 10 shows the distribution of spin current density for the two cases when the width of the output wire is narrower and when it is wider than that of the input wire. Figure 11 shows the magnitude of the spin current for different ratios of the width of the input to output wires. The spin current was calculated by integration of the spin current density over the wire cross-section. Because of the variable width, the distribution along the wire is different for the spin current and for the spin current density.

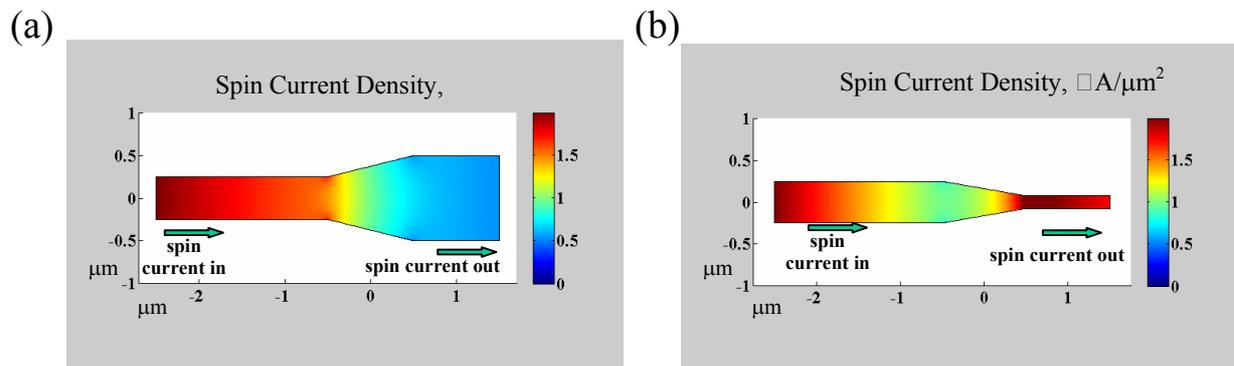

*Fig. 10 Spin current density in wire with variable width. The spin diffusion length in the wire is 5 μm. Input of spin current is from left side. Ratio of width of input wire to width of output wire is : (a) 2; (b) 0.3.*

As in example 2, when a spin drain attracts spin current, the spin relaxation becomes slower. As can be seen from Fig.11, the spin relaxation in the input wire becomes slower in the case of a wider output wire (blue and magenta lines) and the spin relaxation becomes faster in the case of a narrower output wire (green and red lines). Therefore, the wire of wider width could be considered as an effective spin drain.

## 8. Conclusion

Equations, which describe spin and charge transport in materials with spin-dependent conductivity, were derived from the spin and charge conservation laws. It was shown that the spin and charge transport is distinguishably different in the three types of materials: non-magnetic metals, ferromagnetic metals and semiconductors.

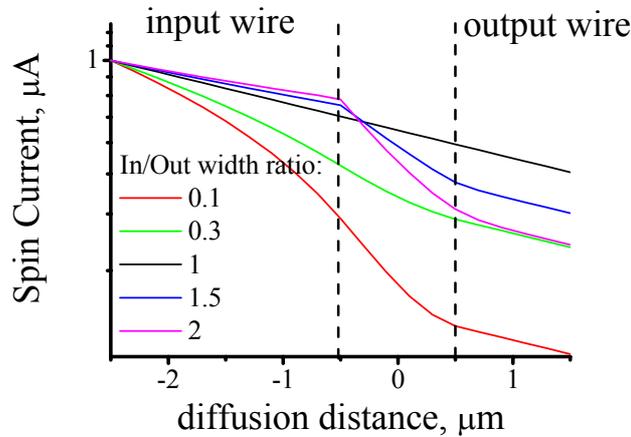

*Fig. 11 Magnitude of spin current along wire with variable width for different width ratios between input to output. Black line represents the case of straight line.*

The conductivity of non-magnetic metals is spin-independent and there is no interaction between the spin diffusion current and the charge drift current in non-magnetic metals. In the case of non-magnetic metals the obtained spin/charge transport equations converge to Valet-Fert spin diffusion equation.

The conductivity of ferromagnetic metals is spin-dependent and the spin selectivity $\beta$ is constant throughout the bulk of the metal. In a ferromagnetic metal the drift current has a spin-current component. In contrast to the diffusion spin current, the drift spin current does not decay along flow direction.

In ferromagnetic metals there is a charge accumulation along the flow of the diffusion spin current. Because of the charge accumulation, the effective diffusion length in ferromagnetic metals becomes shorter. In the case when the spin-selectivity of conductance is near 100 %, the spin diffusion is blocked and the diffusion spin current can not flow. In ferromagnetic metals there is no interaction between the dc diffusion and drift currents, but there is an interaction between the ac diffusion and drift currents.

The conductivity of semiconductors is spin-dependent only in the presence of a spin accumulation. There is a substantial interaction between the diffusion and drift currents in semiconductors. A unique feature of the spin transport in the semiconductors is the existence of the threshold spin current, above which the spins can not diffuse. In the case when magnitude of the diffusion spin current is smaller, but near the magnitude of threshold spin current, a region of high spin and charge accumulation is formed.

In the semiconductors, the drift charge current modulates the effective spin diffusion length of the diffusion current. In the case of the same flow direction of diffusion and drift current, the effective spin diffusion length becomes longer, the spin relaxation becomes slower and the spin diffusion current are gained from the drift current. In the case of the opposite flow directions, the effective spin diffusion length becomes shorter, the spin relaxation becomes faster and the spin diffusion current are damped by the drift current.

A spin drain plays an important role in the spin transport in non-magnetic metal, ferromagnetic metals and semiconductors. The attraction of spin current by the spin drain and the influence of the spin drain on the spin diffusion length always should be considered for design optimization of spintronics devices.

**Appendix 1.**

In the appendix we prove the validity of Eqn. (7), which states that the rate of spin relaxation is linearly proportional to the spin chemical potential $\mu_{spin}$. Also, the features of spin transport in the case, when Eqn. (7) is not valid, are discussed.

The following is the calculation of the spin relaxation for a material in which there is a charge spin accumulation. In the case when the energy is conserved during spin-flip scattering, the rate of spin relaxation is calculated as

$$\frac{\partial n_S}{\partial t} = \int P_\uparrow N_\uparrow(E) \cdot N_\downarrow(E) \cdot F\left(\frac{E - E_F - \Delta E_\uparrow}{kT}\right)\left[1 - F\left(E + \frac{E - E_F - \Delta E_\uparrow}{kT}\right)\right] dE - \\ \int P_\downarrow N_\uparrow(E) \cdot N_\downarrow(E) \cdot F\left(\frac{E - E_F + \Delta E_\downarrow}{kT}\right)\left[1 - F\left(\frac{E - E_F + \Delta E_\downarrow}{kT}\right)\right] dE \quad (A.1)$$

where

$n_S$ is the number of accumulated spins, $P_\uparrow, P_\downarrow$ are the probabilities for the transition from the spin-up band to the spin-down band and transition from the spin-down band to the spin-up band, respectively.
$N_\uparrow, N_\downarrow$ are the densities of states for spin-up to spin-down bands
F(E) is the Fermi-Dirac distribution
$E_F$ is the Fermi energy in case when there is no charge and spin accumulations
$\Delta E_\uparrow, \Delta E_\downarrow$ is the change of the Fermi energy due to a charge and/or spin accumulation.

In the case when there is no spin accumulation, $\Delta E_\uparrow = \Delta E_\downarrow$ and there is no spin relaxation $\frac{\partial n_S}{\partial t} = 0$.
That leads to $P_\uparrow = P_\downarrow = P$.
In the case when the charge and spin accumulations are small

$$\frac{\Delta E_\uparrow}{kT} \ll 1 \quad \frac{\Delta E_\downarrow}{kT} \ll 1 \quad (A.2)$$

The Eqn. (A.1) is simplified to

$$\frac{\partial n_S}{\partial t} = \int P \cdot N_\uparrow(E) \cdot N_\downarrow(E) \cdot \left(F_0 - F'\frac{\Delta E_\uparrow}{kT}\right)\left[1 - F_0 + F'\frac{\Delta E_\uparrow}{kT}\right]dE -$$
$$- \int P \cdot N_\uparrow(E) \cdot N_\downarrow(E) \cdot \left(F_0 + F'\frac{\Delta E_\downarrow}{kT}\right)\left[1 - F_0 - F'\frac{\Delta E_\downarrow}{kT}\right]dE \quad (A.3)$$

Simplifying Eqn. (A.3), we obtain

$$\frac{\partial n_S}{\partial t} = \frac{\Delta E_\uparrow - \Delta E_\downarrow}{kT}\int P \cdot N_\uparrow(E) \cdot N_\downarrow(E) \cdot F'(2F_0 - 1)dE = A \cdot (\Delta E_\uparrow - \Delta E_\downarrow) \quad (A.4)$$

where
$$F_0 = F\left(\frac{E - E_F}{kT}\right) \quad F' = \frac{\partial F\left(\frac{E - E_F}{kT}\right)}{\partial E} \quad A = \frac{1}{kT}\int P \cdot N_\uparrow(E) \cdot N_\downarrow(E) \cdot F'(2F_0 - 1)dE$$

Since
$$\Delta E_\uparrow - \Delta E_\downarrow = \mu_\uparrow - \mu_\downarrow = \mu_{spin} \quad (A.5)$$

Eqn (A.4) will be

$$\frac{\partial n_S}{\partial t} = A \cdot \mu_{spin} \quad (A.6)$$

The expression (A.6) states that if condition (A.2) is satisfied, the spin relaxation is linearly proportional to $\mu_{spin}$. Therefore, the spin diffusion length does not depend on spin or charge accumulation.

In the case when condition (A.2) is not satisfied, the parameter A is a function of the charge accumulation $\rho_{charge}$ and the spin accumulation $\rho_{spin}$.

$$\frac{\partial n_S}{\partial t} = A(\rho_{charge}, \rho_{spin}) \cdot \mu_{spin} \quad (A.7)$$

Since the spin accumulation is proportional to the spin chemical potential and the charge accumulation is proportional to the second derivative of the charge chemical potential, Eqn. (A.7) can be rewritten as

$$\frac{\partial n_S}{\partial t} = A(\mu_{spin}, \nabla^2 \mu_{charge}) \cdot \mu_{spin} \quad (A.8)$$

Therefore, at sufficiently large spin or charge accumulation, the spin diffusion length may depend on the magnitude of the spin or charge accumulation. For example, the dependence of the spin diffusion length on the magnitude of charge and spin accumulations should be considered in the cases when the spin current in a semiconductor is close to the threshold spin current.